\parskip = 2pt
\magnification=\magstep1

\newcount\ftnumber
\def\ft#1{\global\advance\ftnumber by 1
          {\baselineskip=13pt \footnote{$^{\the\ftnumber}$}{#1 }}}
\newcount\fnnumber
\def\fn{\global\advance\fnnumber by 1
         $^{(\the\fnnumber)}$}

\def\fr#1/#2{{\textstyle{#1\over#2}}} 

\font\title = cmbx10 scaled 1440 
\font\ss = cmssbx10

\def\Cc{\hbox{\ss C}}

\def\>{\rangle}
\def\<{\langle}
\def\k#1{|\,#1\>}

\def\x{\otimes}

\def\al{\alpha}
\def\be{\beta}
\def\ga{\gamma}
\def\de{\delta}

\def\Sc{\hbox{\ss S}}

\def\oc{\hbox{\ss 1}}
\def\1c{\hbox{\ss 1}} 
 
\def\0c{\hbox{\ss 0}} 
\def\Pc{\hbox{\ss P}}

\def\Cc{\hbox{\ss C}} 
 
\def\Xc{\hbox{\ss X}}
\def\Yc{\hbox{\ss Y}}
\def\Zc{\hbox{\ss Z}}

\def\Uc{\hbox{\ss U}}
\def\Vc{\hbox{\ss V}}

\def\Cc{\hbox{\ss C}}
\def\Hc{\hbox{\ss H}}

\def\c{\fr1/{\sqrt2}}

\def\x{\otimes}
\def\h{\fr1/2}

\def\ra{\rightarrow}

\def\+{\oplus}

\def\={\equiv}

\newcount\eqnumber

\def\eq(#1){
    \ifx\DRAFT\undefined\def\DRAFT{0}\fi	
    \global\advance\eqnumber by 1%
    \expandafter\xdef\csname !#1\endcsname{\the\eqnumber}%
    \ifnum\number\DRAFT>0%
	\setbox0=\hbox{#1}%
	\wd0=0pt%
	\eqno({\offinterlineskip
	  \vtop{\hbox{\the\eqnumber}\vskip1.5pt\box0}})%
    \else%
	\eqno(\the\eqnumber)%
    \fi%
}
\def\(#1){(\csname !#1\endcsname)}


\baselineskip = 14pt

\font\title = cmbx10 scaled 1440    

\centerline{\title From Cbits to Qbits:}
\medskip
\centerline{\title Teaching Computer Scientists Quantum Mechanics\/}
\bigskip
\centerline{N. David Mermin}
\medskip
\centerline{Laboratory of Atomic and Solid State Physics}
\centerline{Cornell University, Ithaca, NY 14853-2501} 
\bigskip

{\narrower \baselineskip = 12 pt

\bigskip

\noindent  A strategy is suggested for teaching
mathematically literate students, with no background in physics, just
enough quantum mechanics for them to understand and develop algorithms
in quantum computation and quantum information theory.  Although the
article as a whole addresses teachers of physics, well versed in
quantum mechanics, the central pedagogical development is addressed
directly to computer scientists and mathematicians, with only
occasional asides to their teacher.  Physicists uninterested in quantum
pedagogy may be amused (or irritated) by some of the views of standard
quantum mechanics that arise naturally from this unorthodox
perspective.

}

\bigskip

\bigskip
\centerline{{\bf I. Computer science and quantum mechanics}}
\medskip

{\sl \hskip 30pt These ``bras'' and ``kets'' --- they're just vectors!} 

\vskip 6pt
\hskip 150pt   --- Newly enlightened computer scientist\ft{Speaker at second
AQIP (Algorithms in quantum information theory) workshop, Chicago,
January 1999.}

\medskip
There is a new audience for the teaching of quantum mechanics whose
background and needs require a new style of quantum pedagogy.  The
audience consists of computer scientists.  Compared with the usual
students in an introductory quantum mechanics course they are
mathematically sophisticated, but are often ignorant of and
uninterested in physics.  They want to understand the applications of
quantum mechanics during the past dozen years to information
processing, and their focus is exclusively on algorithms
(software), not engineering (hardware).  

Although the obstacles to quantum computers becoming a viable
technology are formidable, the profound consequences of quantum
mechanics for the theory of computation discovered during the past
decade ought to be part of the intellectual equipment of every
computer scientist, if only because it provides dramatic proof that
the abstract analysis of computation cannot be divorced from the
physical means available for its execution.  Computer scientists of
the future ought to learn quantum mechanics.

But how much quantum mechanics?  In December 2001 I was at a
conference on quantum computation and information at the Institute for
Theoretical Physics in Santa Barbara.  At lunch one day I remarked to
the Director of the ITP that I spent the first four or five lectures
of my course\ft{Lecture notes and homework assignments can be found at
{\tt www.ccmr.cornell.edu/ \~{}mermin/qcomp/CS483.html}.  The
pedagogical approach sketched below is fleshed out in Chapter 1.} in
quantum computation teaching the necessary quantum mechanics to the
computer scientists in the class.  His response was that any
application of quantum mechanics that could be taught after only a
four hour introduction to the subject, could not have serious
intellectual content.  After all, he remarked, it takes any physicist
years to develop a feeling for quantum mechanics.

It's a good point.  Nevertheless it is a fact that computer scientists
and mathematicians with no background in physics have been able
quickly to learn enough quantum mechanics to understand and contribute
importantly to the theory of quantum computation, even though quantum
computation repeatedly exploits the most notoriously paradoxical
features of the subject.  There are three main reasons for this:

First, a quantum computer --- or, more accurately, the abstract
quantum computer that one hopes some day to be able to realize --- is
an extremely simple example of a physical system.  It is discrete, not
continuous.  It is made up out of a finite number of units, each of
which is the simplest possible kind of quantum mechanical system, a
2-state system, whose possible behavior is highly constrained and
easily analyzed.  Much of the analytical complexity of learning
quantum mechanics is connected to mastering the description of
continuous (infinite-state) systems in 3+1 dimensional space-time.  By
restricting attention to discrete transformations acting on
collections of 2-state one can avoid much suffering (and lose much
wisdom, none of it --- at least at this stage of the art --- relevant
to the theory of quantum computation.)

Second, the hardest part of learning
quantum mechanics is to get a good feeling for how the abstract
formalism can be applied to actual phenomena in the laboratory.  This
almost invariably involves formulating oversimplified abstract models
of the real phenomena, to which the quantum formalism can effectively be
applied.  The best physicists have an extraordinary intuition for what
features of the actual phenomena are essential and must be represented
in the abstract model, and what features are inessential and can be
ignored.  It takes years to develop such intuition.  Some never do.
The theory of quantum computation, however, is only concerned with
the abstract model --- the easy part of the problem.

Third, to understand how to {\it build\/} a quantum computer, or
to study what physical systems are promising candidates for realizing
such a device, you must indeed have many years of experience in
quantum mechanics and its applications under your belt.  But if you
only want to know what such a device is capable of doing in principle,
then there is no reason to get involved in the really difficult
physics of the subject.\ft{There is also an isolated subset of
quantum-computational theory called adiabatic quantum computation that
requires a somewhat broader view of the quantum theory because it uses
the quantum system more like an analogue than a digital compute.}  The
same holds for ordinary (``classical'') computers: one can be a
masterful practitioner of computer science without having the foggiest
notion of what a transistor is, not to mention how it works.

So while the approach to quantum mechanics for computer scientists
sketched below is focused and limited in scope, it
is neither oversimplified nor incomplete, for the special task for
which it is designed.

\bigskip
\centerline{{\bf II. Classical bits.}}
\medskip

The first step in teaching quantum mechanics to computer scientists is
to reformulate the language of conventional ({\it classical\/})
computation in an unorthodox manner that introduces much of the
quantum formalism in an entirely familiar setting

To begin with, we need a term for a physical system that can exist in
two unambiguously distinguishable states, which are used to represent
0 and 1.  Often such a system is called a {\it bit}, but this can
obscure the important distinction between the abstract bit (0 or 1)
and the physical system used to represent it. If one could establish a
nomenclature for the field at this late date I would argue for the
term Cbit for a classical physical system used to represent a bit, in
parallel with with the term Qbit for its quantum generalization.
Unfortunately the ugly and orthographically preposterous term {\it
qubit\/} currently holds sway for the quantum system, while {\it
bit\/} is used for both the classical system and the abstract bit.  On
the slim chance that it is still not too late for a more sensible
nomenclature, I shall indulge myself by using these two unorthodox
terms.\ft{They are inspired by Paul Dirac's early use of {\it
c-number} and {\it q-number} to describe classical variables and their
generalizations to quantum-mechanical operators.  The prevailing
``qubit'' honors the English rule that $q$ should be followed by $u$
but ignores the equally powerful requirement that $qu$ should be
followed by a vowel.  Although {\it qubit} is pronounced ``cue-bit'',
it's spelling suggests ``k'bit''.  No version of {\it Cbit\/} is in
general use. Cbit and Qbit are preferable to c-bit and q-bit because
the terms themselves often appear in hyphenated constructions:
``multi-Qbit'' is clearly preferable to the ungainly and ambiguous
``multi-q-bit''.}

It can be fruitful, even on the strictly classical level, to represent
the two states of a Cbit by a pair of orthonormal 2-vectors, denoted
by the symbols\ft{This notation for vectors goes back to Dirac.  For
reasons too silly to go into, Dirac called such vectors {\it kets\/},
a terminology that has survived to the present day.}  $$\k0,\
\ \k1.\eq(cbit)$$ To do nontrivial computation requires more than one
Cbit.  It is convenient (and, as we shall see in a moment, even
natural) to represent the four states of two Cbits as four orthogonal
vectors in four dimensions, formed by the tensor products of two
such pairs: $$\k0\x\k0,\ \ \k0\x\k1 \ \
\k1\x\k0\ \ \k1\x\k1.\eq(2cbits)$$  One often omits the $\x$, writing \(2cbits) in the
more compact but equivalent form,
$$\k0\k0\ \ \k0\k1\ \ \k1 \k0 \ 
\ \k1 \k1,\eq(2cbitsa)$$ or, more readably,
$$\k{00}\ \ \k{01}\ 
\ \k{10} \ \ \k{11}, \eq(2qubitsb)$$
or, most compactly of all, using the decimal representation of the
2-bit number represented by the pair of Cbits, 
$$\k{0}_2\ \ \k{1}_2\ \ \k{2}_2 \ 
\ \k{3}_2.\eq(2qubitsc)$$

The subscript 2 is necessary in the last form \(2qubitsc) because in
going from binary to decimal we lose the information of how many Cbits
the vector describes, making it necessary to indicate in some other
way whether $\k3$ means $\k{11} = \k3_2$ or $\k{011} = \k3_3$ or
$\k{0011} = \k3_4,$ etc.

As this last remark illustrates, one represents the states of $n$ Cbits as
the $2^n$ orthonormal vectors in $2^n$ dimensions, 
$$\k x_n, \ \ \ 0 \leq x < 2^n,\eq(ncbit)$$
given by the
$n$-fold tensor products of $n$ mutual orthogonal pairs of orthogonal
2-vectors.

Thus, for example,

$$\k{19}_6 = \k{010011} = \k0\k1\k0\k0\k1\k1 =
                 \k0\x\k1\x\k0\x\k0\x\k1\x\k1.\eq(5cbit)$$

That the tensor product is a convenient and highly appropriate way to
represent multi-Cbit states becomes clear if one expands the vectors
representing each Cbit as column vectors,
$$\k0 \longleftrightarrow   \pmatrix{1\cr0},\ \
\k1 \longleftrightarrow   \pmatrix{0 \cr 1}.\eq(1cbitcol)$$
The corresponding column vectors for tensor products are
$$\pmatrix{y_0\cr y_1}\pmatrix{z_0\cr z_1}
\longleftrightarrow \pmatrix{y_0z_0\cr y_0z_1\cr y_1z_0\cr 
y_1z_1\cr},\eq(2cbitcol)$$
$$\pmatrix{x_0\cr x_1}\pmatrix{y_0\cr y_1}\pmatrix{z_0\cr z_1}
\longleftrightarrow \pmatrix{x_0y_0z_0\cr x_0y_0z_1\cr x_0y_1z_0\cr 
x_0y_1z_1\cr x_1y_0z_0\cr x_1y_0z_1\cr x_1y_1z_0\cr
x_1y_1z_1\cr},\eq(3cbitcol)$$ etc.

Thus, for example, the 8-dimensional column vector representing $\k5_3$
is given by
$$\k5_3 = \k{101} = \k1\k0\k1 = 
\pmatrix{0\cr 1}\pmatrix{1\cr 0}\pmatrix{0\cr 1}
= \pmatrix{0\cr 0\cr 0\cr 0\cr 0\cr 1\cr 0\cr 0\cr}
\matrix{^{_{_0}}\cr^{_{_1}}\cr^{_{_2}}\cr^{_{_3}}\cr^{_{_4}}\cr^{_{_5}}
\cr^{_{_6}}\cr^{_{_7}}}\eq(3cbitcola)$$
which has a 0 in every entry except for a 1 in the entry labeled by
the integer 5 that the three Cbits represent.\ft{Label the entries by
counting down 0,1,2$\ldots$ from the top.  The small numerals on the
extreme right in \(3cbitcola) make this labeling explicit.}  This
general rule for the column vector representing $\k x_n$ --- 1 in
position $x$ and 0 everywhere else --- is the obvious generalization
to $n$ Cbits of the form for a 1-Cbit column vector.  It is an automatic
consequence of standard tensor-product notation.

\bigskip
\centerline{{\bf III. Operations on Cbits.}}
\medskip
 
In quantum computation all operations on Qbits are reversible,\ft{An
example of an irreversible operation is Erase: $\k0
\ra \k0,\ \ \ \k1 \ra \k0$.  It is irreversible because one cannot
reconstruct the input from the output: it has no inverse.} except for
the process called ``measurement'' described in Section VI.
Measurement plays no role in classical computation (or, perhaps more
accurately, a role so trivial that it is not recognized explicitly as
a part of the computational process).  Since Cbit states turn out to
be a (tiny) subset of Qbit states, our reformulation of classical bits
and what can be done with them need only consider reversible
operations on the Cbits.

There are just two reversible operations on a single Cbit:

(1) Do nothing (identity operator $\oc$):
$$\oc\k0 = \k0,\ \ \ \oc\k1 = \k1.\eq(oc)$$

\vskip 10pt   
(2) Flip it (flip operator $\Xc$):
$$\Xc\k0 = \k1,\ \ \ \Xc\k1 = \k0 \hskip 35pt (\sigma_x).\eq(Xc)$$
(I have indicated in parentheses the standard physicists' notation;
quantum computer scientists prefer $\Xc$ to $\sigma_x$.) 

Less trivial reversible operations are available on two Cbits.  One
can, for example, exchange the values of the bits they represent (swap
operator $\Sc$):
$$\Sc\k{xy} = \k{yx}.\eq(exchg)$$ 

In manipulating such muti-Cbit operations it is useful to have a
compact notion for the action on a many-Cbit state of operations that
act on only a single one of the Cbits.  One labels the
Cbits by integers 0,1,2,$\ldots$ (starting with zero on the right)
associated with the power of 2 that each Cbit represents.  Thus if $x$
has the binary expansion $x = 8x_3 + 4x_2 + 2x_1+x_0$, then
$$\k x_4 = \k{x_3x_2x_1x_0} =  \k{ x_3} \k{x_2} \k{x_1} \k{x_0} =
\k{x_3}\x \k{x_2}\x \k{x_1}\x \k{x_0}.\eq(x4bit)$$
An operation that acts only on Cbit \#2 is
$$\Xc_2 = \oc\x\Xc\x\oc\x\oc.\eq(Xc2)$$ Clearly the subscript
indicating which of the four Cbits is subject to the flip operation
$\Xc$ is more transparent than the explicit form of the operator tensor product
on the right.  The subscript notation is unavoidable when large numbers of
Cbits are involved.  It follows from the definition of the operator
tensor product that, as desired,
$$\Xc_2\bigl[\k{x_3}\x\k{x_2}\x\k{x_1}\x\k{x_0}\bigr] =
\k{ x_3}\x \bigl[\Xc\k{x_2}\bigr]\x \k{x_1}\x \k{x_0}.\eq(Xc2a)$$

It is possible to build up meaningful multi-Cbit
operations out of single-Cbit operations that, while formally well
defined, act an on individual Cbit in a way that has no
meaningful classical interpretation.  Here, for example, is a
meaningless operation on one Cbit which can be used to build up
meaningful multi-Cbit operations:
$$\Zc\k0 =\k0,\ \ \Zc\k1 = -\k1 \hskip 25pt (\sigma_z).\eq(Zc)$$

The action of $\Zc$ on the state $\k1$, multiplying it by $-1$, while
mathematically well defined on the 2-dimensional 1-Cbit vector
space, produces a vector that has no meaning within the context of
Cbits.  Only the two vectors $\k0$ and $\k1$ have meaning, as the two
distinguishable states of the Cbit used to represent 0 and 1.  The
whole vector space of linear combinations of the two classical states
is vastly under-utilized, and indeed, the whole use of a $2^n$
dimensional vector space when we are only interested a single set of
$2^n$ orthonormal basis vectors could be viewed as extravagant
conceptual overkill, except for the pleasing structure
introduced by the column-vector representation of the tensor product.
The only classically meaningful reversible operations on $n$ Cbits are
the $(2^n)!$ different permutations of the $2^n$ basis vectors.

Nevertheless a meaningless 1-Cbit operation like $\Zc$ can acquire
classical meaning when used in conjunction with other such meaningless
operations in a multi-Cbit context.  As an important example, notice
that the 2-Cbit operation $\h\bigl(\oc+\Zc_1\Zc_0\bigr)$ acts as the
identity on the 2-Cbit states $\k0\k0$ and $\k1\k1$, while giving 0
(another classically meaningless output) when acting on $\k0\k1$ or
$\k1\k0$.  The operation $\h\bigl(\oc-\Zc_1\Zc_0\bigr)$, on the other
hand, acts as the identity on $\k0\k1$ and $\k1\k0$, while giving 0 on
$\k0\k0$ and $\k1\k1$.  Evidently both are projection operators in the
full vector space spanned by all the 2-Cbit basis states.\ft{More
precisely the projection operators are their linear extensions to the
full space from the basis on which they are defined.  Quite generally
any operation whose action is defined only on the classical basis
states can be identified with its linear extension to the whole vector
space.}  Since the operation $\Sc_{10}$, which exchanges the values of
Cbits $1$ and $0$, acts as the identity if their state is $\k{00}$ or
$\k{11}$ and as the double-flip operator $\Xc_1\Xc_0$ if their state
is $\k{01}$ or $\k{10}$, we are led to the following operator
representation of $\Sc_{10}$:
$$\Sc_{10} =  \h\bigl(\oc+\Zc_1\Zc_0\bigr)
                        +\Xc_1\Xc_0\h\bigl(\oc-\Zc_1\Zc_0\bigr),\eq(Scij)$$
or\ft{Note that
1-Qbit operators acting on different Qbits (like $\Xc_1$ and $\Zc_0$)
commute even though the 1-Qbit operators (\Xc\ and \Zc) do not commute
when acting on the same Qbit.}
$$\Sc_{10} = \h\bigl(\oc+\Zc_1\Zc_0 +\Xc_1\Xc_0 -
\Yc_1\Yc_0\bigr),\eq(Scij1)$$ where
$$\Yc = \Xc\Zc \hskip 30pt (-i\sigma_y).\eq(Yc)$$ I digress to remark
to physicist readers that this ``classical'' derivation of the
exchange operator is much simpler and transparent than the standard
quantum mechanical derivation\ft{One notes that the triplet states of two
spin-$\h$ particles are symmetric, and the singlet state is
antisymmetric, so $\Sc_{10} = \Pc_{\rm triplet} - \Pc_{\rm singlet}$.
Because $s(s+1) = \bigl(\h\sigma^{(1)} + \h\sigma^{(0)}\bigr)^2,$ in
the singlet ($s=0$) state, $0 = \fr3/2 +
\h\sigma^{(1)}\cdot\sigma^{(0)}$, so $\sigma^{(1)}\cdot\sigma^{(0)} =
-3$, while in every triplet ($s=1$) state, $2 = \fr3/2 +
\h\sigma^{(1)}\cdot\sigma^{(0)}$, so $\sigma^{(1)}\cdot\sigma^{(0)} =
1$.  So $\sigma^{(1)}\cdot\sigma^{(0)} = \Pc_{\rm triplet} -
3\Pc_{\rm singlet}$.~{\ }~Since $\oc = \Pc_{\rm triplet} + \Pc_{\rm
singlet}$, adding these last two relations together one recovers
\(Scij1) in the form $\oc + \sigma^{(1)}\cdot\sigma^{(0)} = 2\Pc_{\rm triplet} -
2\Pc_{\rm singlet} = 2\Sc_{10}$, by a considerably more learned but
substantially less direct route.}, which invokes the full-blown theory of
angular momentum.

Another important example of a 2-Cbit
operation is the controlled-NOT or reversible XOR:
$$\Cc_{10}\k x\k y = \bigl(\Xc_0\bigr)^x\k x\k y = \k
x\k{y\+x}\eq(cNOT)$$ (where $\+$ denotes addition modulo 2), which
flips Cbit 0 (the {\it target} Cbit) if and only if Cbit 1 (the {\it
control} Cbit) has the value 1.  We can build this operation out of
1-Qbit projections,
$$\Cc_{10} =
\h\bigl(\oc+\Zc_1\bigr) + \Xc_0\h\bigl(\oc-\Zc_1\bigr) =
\h\bigl(\oc+ \Zc_1 + \Xc_0  - \Xc_0\Zc_1\bigr).\eq(cNOT1)$$
In this form one sees a curious symmetry: interchanging the operations \Xc\ and \Zc\ has the
effect of exchanging the roles of target and control Cbit, converting
$\Cc_{10}$ to $\Cc_{01}$.

A classically meaningless operation that can be used to perform
just this interchange is the {\it Hadamard\/} transform

$$ \Hc = \c(\Xc + \Zc) = \c\pmatrix{1&\phantom{-}1\cr 1&
-1}.\eq(hadamard)$$ This takes the Cbit states $\k0$ and $\k1$ into
the two classically meaningless linear combinations $\h\bigl(\k0 \pm
\k1\bigr)$.  Since
$$\Xc^2 = \Zc^2 = \oc,\ \ \  \Xc\Zc =
-\Zc\Xc,\eq(XZ)$$
it follows that
$$\Hc^2 = \h(\Xc+\Zc)^2 = \oc,\ \ \ \Hc\Xc = (\Xc+\Zc)\Xc =
\Zc(\Xc+\Zc) =
\Zc\Hc,\eq(HXZ)$$
and therefore
$$\Hc\Xc\Hc = \Zc,\ \ \
\Hc\Zc\Hc = \Xc.\eq(HXZ1)$$

Consequently we can use four classically meaningless operations
$\Hc$ to achieve a classically meaningful task: interchanging the role
of target and control Cbits:
$$\Cc_{01} =
\bigl(\Hc_1\Hc_0\bigr)\Cc_{10}\bigl(\Hc_1\Hc_0\bigr)$$

\bigskip
\centerline{{\bf IV. Quantum bits.}}
\medskip

We have represented the states of $n$ Cbits as a basis of $2^n$
orthonormal vectors in a $2^n$-dimensional vector space constructed as
the $n$-fold tensor product of $n$ 2-dimensional vector spaces.  While
the only classically meaningful operations on these vector spaces consist of
permutations of these classical basis vectors, we have been able to construct
such operations, or reveal relations among them, by introducing
classically meaningless operations that multiply basis vectors by
scalars (in particular 0 or $-1$) or (like the Hadamard transformation
\(hadamard)) take them into non-trivial linear combinations.\ft{One such
construction, the form \(Scij1) of the exchange operator, would achieve an
even more pleasing form were we to introduce $\sqrt{-1}$, replacing
$\Yc$ with $i\Yc$.  This would also restore another symmetry, since
$\Xc = \sigma_x$, $i\Yc =
\sigma_y$, and $\Zc = \sigma_z$ are all hermitian.}

One is reminded of arithmetic before the introduction of $\sqrt{-1}$.
By introducing the ``meaningless'' quantity $i$ one is able to
achieve great simplifications among certain relations connecting
purely ``meaningful'' real numbers.  The bold next step is to declare
the meaningless to be just as meaningful, taking full advantage of the
expanded number system.

A major part of quantum mechanics consists of an analogous expansion
of the notion of the state of a Cbit, called in this extended setting
a quantum bit or {\it Qbit\/}.  We democratically expand the set of
meaningful states from the $2^n$ special orthonormal states, known in
this broader setting as the {\it classical basis\/} (or, in the
prevailing but less informative terminology, the {\it computational
basis\/}) to arbitrary unit vectors from the entire vector space
consisting of all linear combinations (called {\it superpositions\/})
of classical basis states with complex coefficients (called {\it
amplitudes\/}).

Thus the general state of a single Qbit is a superposition of the two
classical-basis states
$$\k\psi = \al\k0+\be\k1,\eq(1qbitstate)$$ where the amplitudes $\al$
and $\be$ are complex numbers constrained only by the normalization
condition
$$|\al|^2+|\be|^2=1.\eq(norm)$$ The general state of $n$ Qbits has the
form
$$\k\Psi = \sum_{0\leq x<2^n}\al_x\k x_n,\eq(nqbitstate)$$ with
complex amplitudes constrained only by the normalization condition
$$\sum_{0\leq x<2^n}|\al_x|^2=1.\eq(nnorm)$$

Physics offers many examples of physical systems --- Qbits ---whose
natural description is in terms of states that are precisely these
peculiar generalizations of the states of classical bits that expand
the constrained set of classical basis vectors to the entire complex
vector space that they span.\ft{The most elementary physical examples
are the polarization states of a photon or the spin states of a
spin-$\h$ particle.  For an understanding of quantum-computational
algorithms it is no more important to know about the detailed physics
of such systems than it is to know about the detailed physics of
transistors for an understanding of classical algorithms.}
    
We shall return momentarily to the consequences of a set of Qbits
having such nonclassical states, but the first thing to note is that
by expanding the set of states from the classical basis vectors to
arbitrary unit vectors in the entire space spanned by the classical
basis, we have already introduced one of the most profound differences
between Cbits and Qbits:  

The most general possible state of two Cbits has the form
$$\k\Psi = \k{x_1}\k{x_0}.\eq(gen2Cbit)$$ This can be described as a
state in which Cbit \#1 has the state $\k{x_1}$ and Cbit \#0, the
state $\k{x_0}$: each individual Cbit has a state of its own.  On the
other hand, the most general possible state of two Qbits has the form
$$\eqalign{\k\Psi &= \al_3\k3_2+\al_2\k2_2+\al_1\k1_2+\al_0\k0_2\cr  &=  
\al_3 \k{1}\k{1} +\al_2 \k{1}\k{0} +\al_1 \k{0}\k{1} +\al_0
\k{0}\k{0}.}\eq(gen2Qbit) $$ If each Qbit had a state of its own, this
2-Qbit state would be, under the obvious generalization of the rule for
multi-Cbit states, the tensor product of those two 1-Qbit states.  The
2-Qbit state would thus have the general form
$$\eqalign{\k\psi\k\phi &= \bigl(\al\k1 + \be\k0\bigr) \bigl(\ga\k1 +
\de\k0\bigr)\cr  &=\al\ga\k1\k1 + \al\de\k1\k0 + \be\ga\k0\k1 +
\be\de\k0\k0.}\eq(prod2Qbit)$$ But the state $\k\Psi$ in \(gen2Qbit) cannot have this form unless
$\al_3\al_0 = \al_2\al_1$.  

So in a general multi-Qbit state each individual Qbit has no state of
its own.  This is the first major way in which Qbits differ from
Cbits.

States of $n$ Qbits in which no subset of fewer than $n$ have states
of their own are called {\it entangled\/}.  Generic $n$-Qbit states
are entangled.  The amplitudes in the expansion \(nqbitstate) have to
satisfy special constraints for the state to be a tensor product
of states associated with fewer than $n$ Qbits.

\bigskip
\centerline{{\bf V. Operations on Qbits}}\nobreak\medskip\nobreak
Quantum algorithms are built up out of operations that act linearly on
the state of $n$-Qubits, while preserving the normalization condition
\(nnorm).    The linear
norm-preserving operators on a complex vector space are the {\it
unitary\/} operators.  So the basic ingredients of a quantum
algorithms are unitary operators on the $2^n$-dimensional
complex space:
$$\k\Psi \ra \Uc\k\Psi, \ \ \ \Uc\ {\rm unitary}.\eq(Unitary)$$ The
classical operations\ft{More precisely, their linear extensions, from
the basis on which they are defined, to the whole space.} ---
permutations of the $2^n$ classical basis vectors --- are special
cases of such operators,

The problem of how to implement physically such unitary
transformations is a question of quantum-computational engineering,
just as the question of how to produce permutations of the values of a
collection of Cbits is a question of classical-computational
engineering.  All that need concern the designer of
quantum-computational software, however, is that unitary
transformations constitute the full field of available operations
(except for measurement, as described in Sec. VI.)  For practical
reasons --- software designers should be willing to take into account
constraints suggested by engineering practicalities --- the available
set of unitary transformations is usually restricted to those that can
be built up out of products of unitary transformations, each of which
act only on single Qbits or only on pairs of Qbits, and an important
part of the ingenuity of quantum programming is devoted to how best to
build up more interesting transformations as products of these basic
units.

So if we view the $2^n$ states of $n$ classical bits as the $2^n$
orthonormal basis vectors $\k x_n$ in a $2^n$ dimensional vector
space, and the reversible operations we can perform on the Cbits as
simply the permutations of these basis vectors, then the
generalization to $n$ quantum bits is extremely simple: the states of
Qbits consist of all the normalized complex linear combinations of the
classical basis vectors, and the reversible operations we can perform
on the Qbits consist of all unitary transformations.  The classical
states and classical operations are a very small subset of the quantum
states and quantum operations. 

It looks as if the extension from Cbits to Qbits opens up an
enormously richer landscape of computational possibilities.  While the
state of one Cbit is specified by a single bit of information, to
specify the state of one Qbit requires an infinite amount of
information: two complex numbers constrained only by the normalization
condition \(norm).  And instead of being limited to shuffling a finite
collection of Cbit states through permutation, one can act on Qbits
with a continuous collection of unitary transformations.  Since it is no
more complex a matter to prepare a given state for Qbits than it is
for Cbits and since it is no more complex a matter to implement a broad
range of unitary transformations on Qbits than it is to implement
permutations on Cbits, the extension from Cbits to Qbits would appear
to bring us to a new level of computational power.

{\it But there is a catch!\/} Qbits suffer from a major limitation
which does not afflict Cbits.  Although their state contains vast
amounts of information, given $n$ Qbits in some state $\k\Psi$ there
is nothing you can do to the Qbits that enables you to learn what
$\k\Psi$ is.  There is thus no way to extract anything like the huge
amount of information contained in the amplitudes $\al_x$.  

What, then, are Qbits good for?  How can we exploit their greater
flexibility to do anything useful at all?

\bigskip
\centerline{\bf VI. Measurement: how to squeeze information out of Qbits}
\medskip
\centerline{\sl A. The Born rule.}
\medskip

The very limited possibilities for extracting information is the
second major way in which Qbits differ from Cbits.  If we have $n$
Cbits in the general classical state $\k x_n$ finding out what the
state is --- i.e. learning the number $x$ --- is unproblematic.
Indeed, it is so straightforward that the act of learning the state is
generally not even regarded as a formal part of the computation.  One
simply looks (on a display or a printout).  Importantly, the state of
the Cbits is unaltered by this acquisition of information.  Once
the computer has ceased to operate on the Cbits, their state remains
$\k x_n$ whether or not anybody takes the trouble to ascertain the
particular value of $x$.

Things could not be more different for Qbits.  If one has $n$ Qbits in
the state
$$\k\Psi_n = \sum_x\al_x\k x_n \eq(nPsi)$$ there is nothing one can do
to them to learn the values of the amplitudes $\al_x$.  There is only
one way to extract any information from the Qbits: to {\it measure\/}
them.  Measuring $n$-qubits consists of subjecting them to a device
which produces (at a display or a printout) an integer $x$ in the
range $0
\leq x < 2^n$.  The only link between the state $\k\Psi$ one may have
labored to impose on the Qbits and the value of $x$ revealed by the
measurement is this: the probability of getting the output $x$ is
just $p_x = |\al_x|^2$, where $\al_x$ is the amplitude of $\k x_n$ in
the expansion \(nPsi) of $\k \Psi_n$.\ft{The condition that the states
be unit vectors is thus the condition that the sum of the
probabilities of all the possible measurement outcomes should be 1.}
This is known as the {\it Born rule\/}, after the physicist Max Born.
 
You might think that by measuring repeatedly one could at least get
some good statistics on the distribution of the magnitudes $|\al_x|$
but this possibility of additional partial information about $\k\Psi$
is ruled out by a second fundamental proviso of the Born rule: once
the value $x$ has been indicated by the measurement, the state of the
$n$ Qbits is no longer $\k\Psi_n$, but $\k x_n$.
The postmeasurement
state contains no trace of the information present in the
premeasurement state $\k\Psi$ and is nothing more than the classical
state associated with the value of $x$ indicated by the measuring
device.

Physicists, in a nomenclature that invites misinterpretation, like to
say that the state $\k\Psi_n$ {\it collapses\/} or {\it is reduced\/}
to the state $\k x_n$ by the measurement.  The conservative way to put
it is simply to state the relation between the states immediately
before and immediately after the measurement, in a way that suggests
no mechanism for the change of state, confers no objective status on
it, and makes no commitment to what (if anything) a change in state
implies about what (if anything) has happened to the Qbits themselves.

You might wonder how one can learn anything at all of computational
interest under these wretched conditions.  The general trick is to
produce, through a cunningly constructed unitary transformation, a
superposition
\(nPsi) in which most of the amplitudes $\al_x$ are zero or very 
close to zero, with useful information being carried by any of the
values of $x$ that have a significant probability of being indicated
by the measurement.  It is also important to be seeking information
which, once possessed, can easily be confirmed (e.g. the factors
of a large number) so that one is not misled by the occasional
irrelevant low probability outcome.

Clearly the action of a measurement on the state of $n$ Qubits is
irreversible: any state $\Psi_n$ with  non-zero amplitude $\al_x$ is
capable of becoming the state $\k x_n$ after a measurement.  There is
no way to reconstruct the input from the output.  Measurement is, however, the
only irreversible operation on Qbits. All other operations are unitary.  

The Born rule contains, as a special case, the unproblematic character
of extracting information from Cbits.  If the state $\k\Psi$ of $n$
Qbits happens to be one of the $2^n$ classical-basis states $\k
{x_0}_n$ then $\al_x = 0, x\neq x_0$ and $\al_{x_0}=1$.  So the result
of measuring the Qbits is $x_0$ with probability 1.  The second
proviso of the Born rule then requires that after the measurement the state
of the Qbits is $\k{x_0}_n$ --- i.e. the post-measurement state
continues to be what it was before the measurement.  The statistical,
state-altering character of the outcome of a measurement of $n$ Qbits
in a general state becomes the deterministic, state-preserving,
unproblematic classical extraction of information when the state is
one of the $2^n$ classical states.

A technical remark for physicists: In this approach to quantum
mechanics it is useful to restrict the term ``measurement'' to
what a broader and more conventional use of the term would
characterize as ``measurement in the classical basis.''  Since
measurement in any other basis can be accomplished by applying an
appropriate unitary transformation --- one that takes the basis of
interest into the classical basis --- followed by measurement in the
classical basis, this restriction of the scope of the term
``measurement'' does not preclude any of the more general possibilities.

\medskip
\centerline{\sl B. Generalization of the Born rule to partial
measurements.}\nobreak\medskip\nobreak There is a generalization of
the Born rule, not often explicitly noted in quantum-mechanics texts,
that often arises in quantum computations.  This happens when only
some of the Qbits are measured.  Suppose we have $m+n$ Qbits and we
decide to measure only $m$ of them.  By representing the $m+n$ bit
number $z$ as $x,y$, the concatenation of the $m$ and $n$ bit binary
strings representing $x$ and $y$, we can write the state of the $m+n$
Qbits as
$$\k\Psi_{m+n} = \sum_{x,y}\al_{x,y}\k{x,y}_{m+n}.\eq(Psimn)$$ Suppose
we we decide to measure only the $m$ Qbits on the left.\ft{Although we
take the $m$ Qbits on the left as the ones to be measured, the rule
for the more general case is the obvious generalization of the one
enunciated below.}  The generalized Born rule states that the
measurement will indicate $x$,\ $0\leq x <2^m,$ with probability
$$p_x = \sum_{0 \leq y < 2^n}|\al_{x,y}|^2,\eq(pxm)$$ and that after
the value of $x$ is indicated, the state of the $m+n$ Qbits changes from
$\k\Psi_{m+n}$ to $\k x_m\k{\Phi_x}_n$, where$$\k{\Phi_x}_n =
p_x^{-1/2}\sum_y\al_{x,y}\k y_n.\eq(genBornPsi)$$

If one immediately follows a measurement of the $m$ Qbits on the left,
with a measurement of the remaining $n$ Qbits on
the right, then this ought to be tantamount to directly measuring all
$m+n$ Qbits.  And indeed, if one applies the generalized Born rule
twice --- first to the measurement of the $m$ Qbits on the left and
then to the measurement of the remaining $n$ on the right --- one does
indeed recover the ordinary Born rule.  

Although the generalized Born rule does not follow from the ordinary
Born rule, it is equivalent to the ordinary Born rule supplemented by
two very reasonable further conditions:

(1) Suppose that between time $t$ and $t'$ no unitary transformations
act on the $m$ Qbits on the left, but arbitrary unitary
transformations may act on the $n$ Qbits on the right --- i.e. the
only unitary transformations acting on the $m+n$ Qbits between $t$ and
$t'$ are of the form $\Uc = \oc_m\x\Vc_n$.  Then the statistical
distribution of outcomes if all $m+n$ Qbits are measured at time $t'$
is unaltered if the $m$ Qbits on the left are measured at any earlier
time between $t$ and $t'$.  Informally, once the computer ceases from
further action on any group of Qbits, you do not have to wait to
the end of the full computation before measuring those Qbits.
 
(2) For a group of $n$ Qbits to be in the state $\k\Phi$ means nothing
more (or less) than that if the Qbits are measured after the application of
an arbitrary unitary transformation $\Vc$, then the distribution of
measurement outcomes will be that specified by the Born rule for $n$
Qbits in the state $\Vc\k\Phi$.

\bigskip

The most important principles formulated in Sections II-VI are
in the table that follows, which summarizes the relevant features of
Qbits by contrasting them to the analogous features of Cbits.
In the table I have introduced the term ``Bit'' (with an upper-case
$B$) to mean ``Qbit or Cbit'' (as opposed to ``bit'' (with a
lower-case $b$, which meas ``0 or 1'').

\bigskip
\def\hs{\hskip 5pt}
\centerline{
\setbox\strutbox=\hbox{\vrule height18pt depth15pt width0pt}
\vbox{\offinterlineskip
\hrule
\halign{\strut\vrule\hfil #\thinspace\hfil 
        &\vrule\vrule\vrule\vrule\hfil #\thinspace\hfil
	&\vrule\hfil # \hfil\vrule\cr 
\noalign{\hrule}
{\bf\   CLASSICAL vs.~QUANTUM BITS\ }  & {\bf Cbits} & {\bf Qbits} \cr
\noalign{\hrule}
\noalign{\hrule}
\noalign{\hrule}
\noalign{\hrule}
\noalign{\hrule}
{\bf States of $n$ Bits}   
& $\ \k x_n,\ \  0 \leq x < 2^n$ & $\sum \alpha_x\k x_n,\ \  
                                    \sum|\alpha_x|^2 = 1$ \cr
\noalign{\hrule}
{\bf Subsets of $n$ Bits} &\ Always have states
                          & Generally have no states \cr
\noalign{\hrule}
\hs {\bf Reversible operations on states}\hs  & \hs   Permutations \hs 
                            & \hs Unitary transformations \hs\cr
\noalign{\hrule}
 {\bf\ Can state be learnt from  Bits?\ }   & Yes    & No \cr
\noalign{\hrule}
{\bf To get information from  Bits} & Just look   & Measure \cr
\noalign{\hrule}
{\bf Information acquired} &   $x$  &  $x$ with probability 
                                             $|\alpha_x|^2$ \cr
\noalign{\hrule}
{\bf\ State after information acquired\ } &\ Same: still $\k x$
                          & Different: now $\k x$ \cr
\noalign{\hrule}
\noalign{\hrule}
\noalign{\hrule}
\noalign{\hrule}
\noalign{\hrule}}}}

\bigskip
\centerline{{\bf VII.  Cautionary remarks and quasi-philosophical reflections}}

\medskip
\centerline{\sl A. An important warning}
\medskip
It is extremely important to avoid a very tempting misinterpretation
--- a gross oversimplification --- of quantum superpositions of
classical states, as illustrated by the following simple example:

A Qbit in the state $\k\psi = \c\k0+\c\k1$ is {\it not\/} the same as
as a Qbit that is either in state $\k0$ or state $\k1$ with equal
probability, even though in either case a measurement will indicate 0
or 1 with equal probability.  To see that the two cases are inherently
different, suppose a Hadamard transformation $\Hc =
\c\bigl(\Xc+\Zc\bigr)$ is applied to the Qbit just before the
measurement is made.  Since
$$\Hc\k0 = \c\bigl(\k0+\k1),\ \ \ \Hc\k1 = \c\bigl(\k0-\k1),\eq(H01)$$
in the second case, whether the initial state is $\k0$ or $\k1$, the
measurement after \Hc\ is applied will continue to indicate 0 or 1
with equal probability.  But in the first case, in which the initial
state is $\k\psi =
\c\bigl(\k0+\k1\bigr)$, we have $\Hc\k\psi = \k0$ so the
measurement after \Hc\ is applied must necessarily indicate 0.

A Qbit in a superposition of classical-basis states is distinctly
different from a Qbit that is in one of those classical states with a
probability given by the squared modulus of the corresponding
amplitude.  Superpositions have no classical interpretation.  They are
{\it sui generis\/}, an intrinsically quantum-mechanical construct,
whose meaning derives only from the rules that characterize the
reversible operations (unitary) that can be performed on them and
the available means (measurement) for extracting information from them.

\medskip
\centerline{\sl B. Meaning of the quantum state}
\medskip

People have been arguing about the meaning of the quantum state ever
since the concept first appeared, with no indication that we
are getting any closer to a consensus.  These conceptual issues are
unimportant for an understanding of quantum computation which only requires
one to know how states are built up from other states (by
appropriate unitary transformations) and how information can be
extracted from Qbits in a given state (by measurement, according to
the Born rules).

The initial state on which the unitary transformations operate is
usually a classical-basis state $\k x_n$.  Such a state can be
unambiguously identified as the post-measurement state of $n$ Qbits
after a measurement that indicated the value $x$.  From this point of
view the computational process begins and ends with a measurement, and
the entire role of the state of the Qbits at any stage of a succession
of unitary transformations is to encapsulate the probability of the
outcomes, should the final measurement be made at that stage of the
process, or to enable one to calculate outcome probabilities should
further unitary transformations be applied before the measurement.

The notion that the state of $n$ Qbits is simply a convenient compact
mathematical device for calculating the correlations between the
outcomes of two measurements on those Qbits, between which an
arbitrary unitary transformation may have been applied, is often
associated with the constellation of ideas about quantum mechanics
called the {\it Copenhagen interpretation}.  It is to be contrasted
with the notion that the state of $n$ Qbits is an objective physical
property of those Qbits, in the same strong sense that we can view the
state of $n$ Cbits --- the unique value $x$ that they represent --- as
an objective property of those Cbits.  People who regard the quantum
state as objective in this sense tend to make a fuss about the fact
that there are two quite different ways in which Qbits can change:
deterministically and continuously (if one builds each unitary
transformation out of many infinitessimal ones) via unitary
transformations, and statistically and discontinuously via
measurements.  This dichotomy looses its content if one replaces
``Qbits'' by ``the state of Qbits'' and recognizes that the state is
nothing but a catalog of how different unitary transformations will
result in different distributions of measurement outcomes ---
classical basis states, which alone can be viewed as objective.

Another pitfall of taking their state to be an objective property of
the Qbits is that one can then succumb to the temptation to believe
that the application of a series of unitary transformations to the
Qbits implements a physical computation of all the resulting
amplitudes $\al_x$.  The clue that this has not been accomplished lies
in the fact, noted above, that given the Qbits there is nothing
whatever you can do with them to reveal the values of those
amplitudes.

There are nevertheless some who believe that all the amplitudes
$\al_x$ have acquired the status of objective physical quantities,
inaccessible though those quantities may be.  Such people then wonder
how that vast number of high-precision calculations ($10^{30}$
different amplitudes if you have 100 Qbits) could all have been
physically implemented.  Those who ask such questions like to provide
sensational but fundamentally silly answers involving vast numbers of
parallel universes, invoking a point of view known as the {\it many
worlds\/} interpretation of quantum mechanics.  My own opinion is
that, imaginative as this vision may appear, it is symptomatic of a
lack of a much more subtle kind of imagination, which can grasp
the exquisite distinction between quantum states and objective
physical properties that quantum physics has forced upon us.

\medskip
\centerline{C. Where's Planck's constant?}

\vskip 6pt

\hskip 30pt  {\sl Where's h-bar?} {\it Where is h-bar?!}

\hskip 150pt   --- Disgruntled quantum optician.\ft{Private
communication to the author at the International Conference on Quantum
Information, Rochester, June 2001.}
\medskip
Like my disapproving colleague, some physicists may be appalled to have
finished what purports to be an exposition of quantum mechanics ---
indeed, of applied (well, {\it gedanken\/} applied) quantum mechanics
--- without ever having run into Planck's constant.  How can this
be?

The answer goes back to my first reason why enough quantum mechanics
to understand quantum computation can be taught in a mere four hours.
We are interested in discrete (2-state) systems and discrete (unitary)
transformations.  But Planck's constant only appears in the context of
continuously infinite systems (position eigenstates) and continuous
families of transformations (time development) that act on them.  Its
role is to relate the conventional units in which we measure space and
time, to the units in which it is quantum-mechanically natural to take
the generators of the unitary transformations that produce
translations in space or time.

If we are not interested in location in continuous space and are only
interested in global rather than infinitessimal unitary
transformations, then $\hbar$ need never enter the story.  The
engineer, who must figure out how to implement unitary transformations
acting over time on Qbits located in different regions of physical
space, must indeed deal with $\hbar$ and with Hamiltonians that
generate the unitary transformations out of which the computation is
built.  But the designer of algorithms for the finished machine need
only deal with the resulting unitary transformations, from which
$\hbar$ has disappeared as a result, for example, of judicious choices
by the engineers of the times over which the interactions that produce
the unitary transformations act.  

Deploring the absence of $\hbar$ from expositions of quantum computer
science is rather like complaining that the $I$-$V$ curve for a
$p$-$n$ junction never appears in expositions of classical computer
science. It is to confuse computer {\it science\/} with computer 
{\it engineering}.
 
\bigskip
\centerline{\bf VIII. That is all you need to know}
\medskip

Armed with the contents of Sections II-VI, one is ready to embark on
the exposition of quantum computer science.  To be sure, there will be
times when it is convenient to expand upon the minimal formalism
developed above.  But such expansions, for example the introduction of
{\it bras} (as linear functionals on the kets), the introduction of
density matrices, or the useful connection between \Xc,
\Yc, and
\Zc\ and the group of 3-dimensional rotations, are all
technical mathematical refinements within the basic structure of the
complex vector space of Qbits.  They require no new physical
principles for their development.

Sections II-VI provide all the quantum mechanics one needs to develop
fully the factorization algorithm of Peter Shor, the search
algorithm of Lov Grover, and their later generalizations.  Only in
developing the very important subject of quantum error correction is
it necessary to introduce a new physical assumption, that the formalism
developed to describe Qbits --- quantum states, unitary
transformations, the Born rules --- describes not only Qbits, but
anything else in the world that the Qbits might happen to interact
with.

If this far from modest extension of the scope of the formalism proves
too big a pill for computer scientists to swallow, one can compromise
with a more limited model of error correction, in which the computer
contains large numbers of extraneous Qbits.  Ideally, these irrelevant
Qbits are perfectly uncoupled to the Qbits of interest, in the sense
that all unitary transformations act only on the Qbits of interest or
only (unimportantly and uninterestingly) on the extraneous Qbits.  But
unfortunately there is a small amount of unintended coupling between
the two sets of Qbits --- unitary transformations whose action is not
restricted to either the relevant or irrelevant Qbits --- whose
disruptive action on the relevant Qbits it is the task of error
correction to undo.  One can then remark, as an aside, that parts of
the world outside the computer (or computationally irrelevant internal
degrees of the computer), that cannot be be perfectly isolated from
the parts that do the computation, can always be well modeled as just
such collections of extraneous Qbits.

A detailed view of how to erect the edifice of quantum computation on
this foundation can be found in Chapters 2-5 of Ref.~2.  Chapter 6
describes a few further topics in the broader area of quantum
information that can be built on this same foundation.

\bigskip

\noindent {\sl Acknowledgment.}  Supported by the National Science 
Foundation, Grant No.~PHY0098429.
\vfil\eject

\bigskip
\vfil\eject

\bye